\documentclass[%
 aip,
 apl,%
 amsmath,amssymb,
reprint,%
]{revtex4-1}
\usepackage{graphicx}
\usepackage{dcolumn}
\usepackage{bm}

\begin{document}

\preprint{AIP/APL}

\title {Enhancing intrinsic detection efficiency of superconducting nanowire single-photon detectors via helium ion irradiation}

\author{W. J. Zhang}
\affiliation{
State Key Lab of Functional Materials for Informatics, Shanghai Institute of Microsystem and Information Technology (SIMIT), Chinese Academy of Sciences (CAS), 865 Changning Rd., Shanghai, 200050, China}
\affiliation{Center for Excellence in Superconducting Electronics (CENSE), Chinese Academy of Sciences (CAS), 865 Changning Rd., Shanghai, 200050, China}

\author{Q. Jia}%

\affiliation{
State Key Lab of Functional Materials for Informatics, Shanghai Institute of Microsystem and Information Technology (SIMIT), Chinese Academy of Sciences (CAS), 865 Changning Rd., Shanghai, 200050, China
}%

\author{L. X. You}
 \email{lxyou@mail.sim.ac.cn}
\affiliation{
State Key Lab of Functional Materials for Informatics, Shanghai Institute of Microsystem and Information Technology (SIMIT), Chinese Academy of Sciences (CAS), 865 Changning Rd., Shanghai, 200050, China
}%
\affiliation{Center for Excellence in Superconducting Electronics (CENSE), Chinese Academy of Sciences (CAS), 865 Changning Rd., Shanghai, 200050, China}

\author{X. Ou}
 \email{ouxin@mail.sim.ac.cn}
\affiliation{
State Key Lab of Functional Materials for Informatics, Shanghai Institute of Microsystem and Information Technology (SIMIT), Chinese Academy of Sciences (CAS), 865 Changning Rd., Shanghai, 200050, China
}%

\author{H. Li}
\affiliation{
State Key Lab of Functional Materials for Informatics, Shanghai Institute of Microsystem and Information Technology (SIMIT), Chinese Academy of Sciences (CAS), 865 Changning Rd., Shanghai, 200050, China
}%
\affiliation{Center for Excellence in Superconducting Electronics (CENSE), Chinese Academy of Sciences (CAS), 865 Changning Rd., Shanghai, 200050, China}

\author{L. Zhang}
\affiliation{
State Key Lab of Functional Materials for Informatics, Shanghai Institute of Microsystem and Information Technology (SIMIT), Chinese Academy of Sciences (CAS), 865 Changning Rd., Shanghai, 200050, China
}%
\affiliation{Center for Excellence in Superconducting Electronics (CENSE), Chinese Academy of Sciences (CAS), 865 Changning Rd., Shanghai, 200050, China}

\author{Z. Wang}
\affiliation{
State Key Lab of Functional Materials for Informatics, Shanghai Institute of Microsystem and Information Technology (SIMIT), Chinese Academy of Sciences (CAS), 865 Changning Rd., Shanghai, 200050, China
}%
\affiliation{Center for Excellence in Superconducting Electronics (CENSE), Chinese Academy of Sciences (CAS), 865 Changning Rd., Shanghai, 200050, China}

\author{X. M. Xie}
\affiliation{
State Key Lab of Functional Materials for Informatics, Shanghai Institute of Microsystem and Information Technology (SIMIT), Chinese Academy of Sciences (CAS), 865 Changning Rd., Shanghai, 200050, China
}%
\affiliation{Center for Excellence in Superconducting Electronics (CENSE), Chinese Academy of Sciences (CAS), 865 Changning Rd., Shanghai, 200050, China}


\begin{abstract}
Realizing an NbN superconducting nanowire single-photon detector (SNSPD) with a 100\% intrinsic detection efficiency (IDE) at the near-infrared wavelengths is still challenging. Herein, we developed a post-processing method to increase the IDE of NbN SNSPDs to near unity using a 20 keV helium ion irradiation. The IDE enhancement was achieved owing to the ion-induced reduction of the superconducting energy gap and the electron density of states at the Fermi level, determined with the electrical and magnetic transport measurements. The change in optical absorptance of the irradiated SNSPD was negligible as confirmed by the measured optical reflectance and system detection efficiency (SDE). Benefited with the IDE enhancement, the SDE of an irradiated device was significantly increased from 49\% to 92\% at 2.2 K for a 1550 nm wavelength.
%
\end{abstract}

\maketitle

%


Controllable modification of the physical properties of thin films is crucial for their application in different devices. Ion irradiation, which creates defects with well-controlled density and topology is one of the powerful tools for tuning the properties of semiconductors\cite{ref1,ref2} and superconductors\cite{ref3,ref4,ref5}. Studies on the ion irradiation effects have attracted great interest from both science and application, especially for superconductors. Because they can be used as a phase-sensitive method to both understand the superconductivity\cite{ref6} and tune the performance of superconducting devices\cite{ref7}. The ion irradiation effects depend on the mass and energy of the irradiating ions as well as the type of superconductors\cite{ref3,ref4,ref5,ref8}. 
Previously, the He ion irradiation on NbN films which induced increasing of vacancies in both the Nb and N sublattices and then reduced the electron density of states (\textit{ N}${}_{0}$) at the Fermi level has been reported\cite{ref3}.

Superconducting nanowire single-photon detectors (SNSPDs), which have demonstrated unparalleled performance in near-infrared photon detection with high system detection efficiency (SDE, $\mathrm{>}$90\%)\cite{ref9,ref10}, low dark count rate (DCR, $\mathrm{<}$1 Hz)\cite{ref11}, and high temporal resolution ($\mathrm{<}$15 ps)\cite{ref12}, are successfully employed in quantum information processing\cite{ref13,ref14}, and high-speed optical communication\cite{ref15}. 
To date, SNSPDs are often fabricated from 5--8 nm thick NbN films, forming 50--100 nm wide nanowires. The NbN SNSPD operates at a temperature range of 2--4 K with a commercial Gifford-McMahon (GM) cryocooler\cite{ref10,ref16}. However, achieving a saturated intrinsic detection efficiency (IDE) for the near-infrared photons with an NbN SNSPD is difficult because of its relatively high critical temperature (\textit{T}${}_{c}$) or superconducting energy gap (\textit{$\mathit{\Delta}$}) with respect to those of low gap materials\cite{ref18,ref19,ref20,ref21}. Attempts to tune the performance of superconducting devices by varying their chemical components have been made by several groups\cite{ref22}. However, there are few reports on the post-processing that could directly control and compare the performance of SNSPDs.

The IDE indicates the probability of a pulse generation in the nanowire when a photon is absorbed. The detection mechanism of SNSPD relies on the conversion of the energy of the absorbed photon into elementary excitations in the superconducting nanowire. Depending on the applied bias current (\textit{I}${}_{b}$) compared to the depairing critical current \textit{I}${}_{c}$${}_{d}$, the minimum energy \textit{E}${}_{min}$ (or maximum wavelength) detectable can be determined using the modified hotspot model\cite{ref23}:
\begin{equation} \label{1}
E_{min}=hv=\ \frac{hc}{\lambda }\ \ge \ \frac{N_0{\Delta }^2wd}{\zeta }\ \sqrt{\pi D{\tau }_{th}}\ \left(1-\frac{I_b}{I_{cd}}\right),
\end{equation}
where \textit{c} is the speed of light, \textit{$\lambda$} is the photon wavelength, \textit{wd} is the cross-sectional area of the nanowire, \textit{D} is the electronic diffusion coefficient, and \textit{$\tau$${}_{th}$} is the time scale of the quasi-particle multiplication process. From Eq.~\eqref{1}, the IDE of SNSPDs can be improved by fabricating devices with thin and narrow nanowires\cite{ref24}, or with low \textit{$\mathit{\Delta}$} or low \textit{N}${}_{0\ }$materials\cite{ref9,ref18,ref19,ref20}. However, both the thickness and the width of the superconducting nanowire influence \textit{$\mathit{\Delta}$} and optical absorptance as well\cite{ref10,ref16}. As a result, optimizing such parameters simultaneously is technically challenging. Developing a method to tune the properties of SNSPDs without changing the geometric parameters and the optical absorptance of the nanowire would greatly contribute to the design and fabrication of SNSPDs.

Here, we report a novel method to tune the physical properties of superconducting NbN ultrathin films and SNSPDs using He ion irradiation. We first studied the modification of polycrystalline NbN films by different He ion fluences (\textbf{\textit{F}${}_{i}$}) ranging from 1 $\times$ 10${}^{14}$ to 5 $\times$ 10${}^{16}$ ion cm${}^{\mathrm{-}}$${}^{2}$ at an ion energy of 20 keV. Both \textit{$\mathit{\Delta}$} and\textit{ N${}_{0}$} were decreased by the ion irradiation, and the optical absorptance of the irradiated film remained essentially unchanged. Then, the SNSPDs were He ion-irradiated, and their IDE was significantly improved. With this technology, the SDE of an SNSPD was increased from 49\% to 92\% at 2.2 K, and over 90\% at 2.5 K, for the 1550 nm.

The polycrystalline NbN film was deposited on a distributed Bragg reflector (DBR) substrate by direct current reactive magnetron sputtering in a mixed Ar and N${}_{2}$ atmosphere at room temperature, with thickness controlled by the deposition time and verified by transmission electron microscopy (TEM). The DBR made of fifteen SiO${}_{2}$/Ta${}_{2}$O${}_{5}$ bi-layers was used for increasing the optical absorptance of the nanowires to near unity\cite{ref10}. For the electrical and magnetic transport measurements, a four-terminal micro-bridge (40 $\mu$m long, 60 $\mu$m wide) was patterned using ultraviolet lithography and etched by reactive ion etching (RIE). To measure the optical reflectance (\textit{r}) as a function of wavelengths, a spectrophotometer with a scanning step of 1 nm and a relative uncertainty below 0.2\% was used.

\begin{figure}
\includegraphics[width=3.7in]{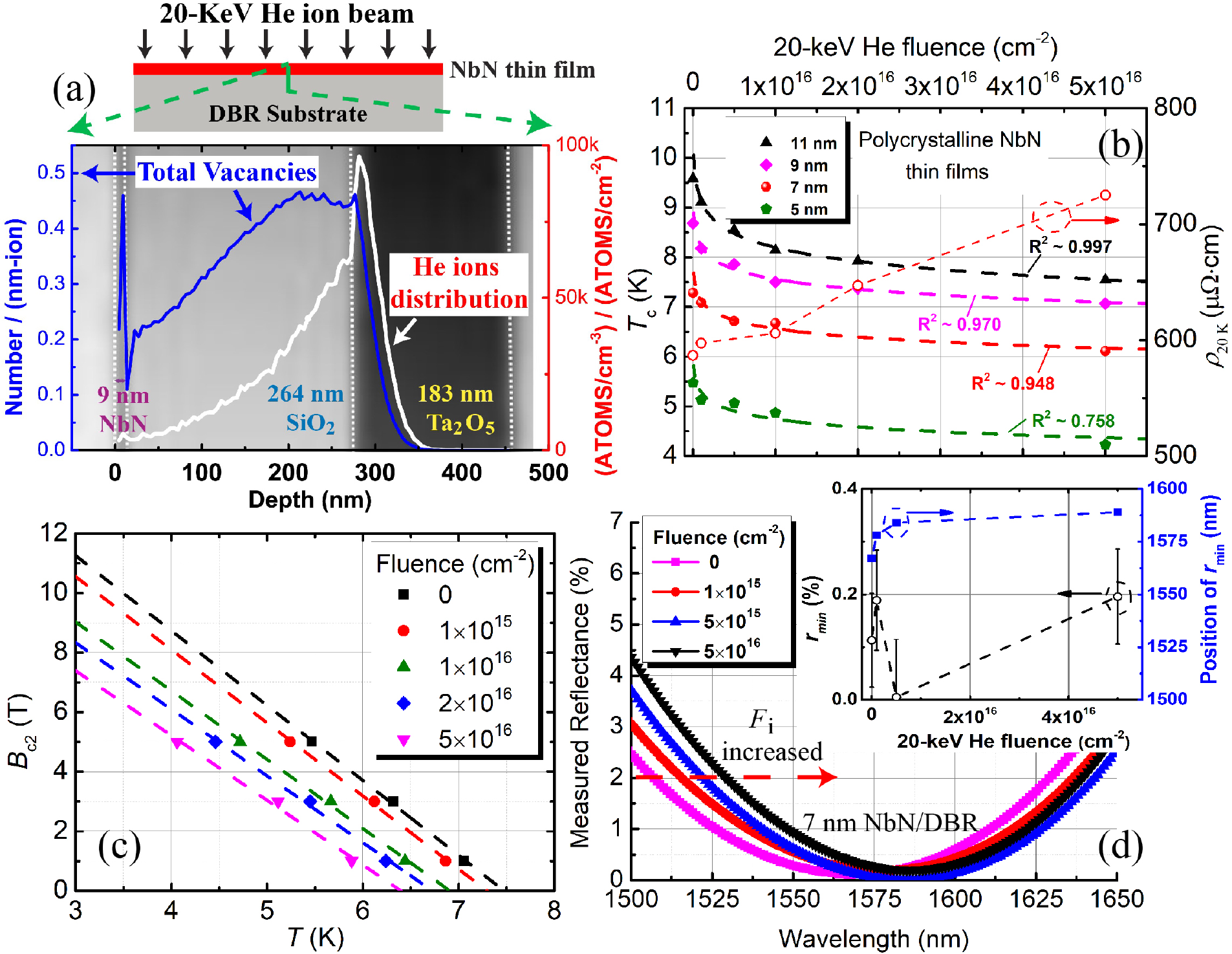}
\caption{\label{fig:fig1}(a) Top: Schematic of the He ion${}^{\ }$irradiation of the NbN thin film on a DBR substrate. Bottom: Simulated vacancy and ion distributions as a function of depth with SRIM method, embedded with a TEM photo of a 9 nm thick NbN film on DBR substrate. (b) Left axis: \textit{F}${}_{i}$ dependence of \textit{T}${}_{c}$ of NbN films with different thicknesses. The solid scatters and the dashed lines represent the experimental data and the logarithmic fittings, respectively. Right axis: \textit{F}${}_{i}$ dependence of the resistivity (\textit{$\rho$}$_{20 K}$, open circles) of the 7 nm thick NbN film. (c) Temperature dependence of \textit{B}${}_{c2}$ for the 7 nm thick NbN film at different \textit{F}${}_{i}$s. Experimental data and fitting results are indicated by solid scatters and dashed lines, respectively. (d) Wavelength dependence of the reflectance (\textit{r}) for the 7 nm thick NbN film on the DBR substrate at different \textit{F}${}_{i}$s. Inset: \textit{F}${}_{i}$ dependence of \textit{r}${}_{min}$ and its positions (solid squares) .
}
\end{figure}

Three SNSPDs (named d1, d2, and d3) with typical meander geometry were fabricated from NbN thin films on DBR substrates by electron beam lithography and RIE\cite{ref10}. These devices were all fabricated in the same run. To guarantee a reliable optical coupling, their active area was designed to be 18  $\times$ 18 $\mu$m${}^{2}$. In this study, a low-energy He ion irradiation with an ion energy of 20 keV and a substrate temperature of about 293 K was used. The choice of He ions is to prevent the chemical reactions and the ion etching effect. The irradiation fluence ranged between 1 $\times$ 10${}^{14}$ and 5 $\times$ 10${}^{16}$ ion cm${}^{\mathrm{-}}$${}^{2}$. Figure~\ref{fig:fig1}(a) shows a schematic layout of the ion irradiation process (top panel) and the distributions of He ions and irradiation induced vacancies for a 9 nm thick NbN film deposited on the DBR substrate (bottom panel). The He ion and vacancy profile were simulated by Monte Carlo simulation with the computer code of the stopping and range of ions in matter (SRIM)\cite{ref25}. The simulation indicated that the project range of He ions was located at the underlying SiO${}_{2}$/Ta${}_{2}$O${}_{5}$ interface, and the top NbN layer was modified by vacancies rather than He ions.

To characterize the optical--electrical properties of the SNSPD, the device was first packaged with a lens single-mode fiber and then cooled with a 2.2 K GM cryostat, as reported previously\cite{ref10}. The input photon flux emitted from a 1550 nm fs laser was attenuated with a serial tunable attenuator to a level of about 1 $\times$ 10${}^{6}$ photon s${}^{\mathrm{-}}$${}^{1}$, calibrated with a high precise optical power meter. The SDE at 1550 measured in this study, with a relative uncertainty of about 3\%\cite{ref10}, was referred to the photons polarized parallel to the nanowire. The electrical and magnetic transport measurements were performed using a Quantum Design PPMS-9 with a temperature stability of $\pm3$ mK and a magnetic field perpendicular to the DBR substrate. 
The resistivity (\textit{$\rho$}) of the samples was recorded as a function of the rising temperature. \textit{T}${}_{c}$ was determined from the temperature at which \textit{$\rho$}(\textit{T}) drops to 50\% of its normal state value at 20 K (\textit{$\rho$}${}_{20 K}$).


\begin{table*}
\caption{\label{tab:TABLE I}Physical parameters of the 7 nm NbN thin film on a DBR substrate and the SNSPD d3 at different \textit{F}${}_{i}$s. 
 }
\begin{ruledtabular}
\begin{tabular}{m{0.7in}m{0.4in}m{0.4in}m{0.4in}m{0.4in}m{0.3in}m{0.4in}m{0.4in}m{0.6in}m{0.4in}m{1in}} 
\textbf{\textit{}} & \centering{\textbf{\textit{F}${}_{i}$\newline (cm${}^{-}$${}^{2})$}}\textbf{\textit{}} & \centering{\textbf{\textit{$\boldsymbol{\rho}_{300K}$}\newline($\mu\Omega\cdot$cm)}}
& \centering{\textbf{\textit{$\boldsymbol{\rho}_{20K}$}\newline($\mu\Omega\cdot$cm)}}
&\centering {\textbf{RRR}} &\centering \textbf{\textit{T${}_{c}$(0)}\newline (K)} & \centering \textbf{\textit{$\boldsymbol{\mathit{\Delta}}$T${}_{c}$(0)}\newline (K)} & \textbf{\centering\textit{$\boldsymbol{\xi}$${}_{GL}$(0)}\newline (nm)} & \centering{\textbf{\textit{D}\newline(cm${}^{2}$ s${}^{-1}$)} }& \centering \textbf{\textit{$\boldsymbol{\mathit{\Delta}}$(0)}\newline (meV)} & \textbf{\centering {\textit{N${}_{0}$}\newline (nm${}^{-}$${}^{3}$ eV${}^{-}$${}^{1}$)}} \\ 
\textbf{\centering{7 nm\newline NbN film\textit{}}}& \textbf{0} & 462 & 587 & 0.787 & 7.28 & 1.55 & 5.1 & 0.44 & 1.11 & 24.4 \\ 
\textbf{\textit{}} & \textbf{1$\times$10${}^{15}$} & 468 & 597 & 0.784 & 7.09 & 1.55 & 5.2 & 0.45 & 1.08 & 23.4 \\
\textbf{\textit{}} & \textbf{1$\times$10${}^{16}$} & 471 & 606 & 0.777 & 6.67 & 1.50 & 5.6 & 0.47 & 1.01 & 21.7 \\ 
\textbf{\textit{}} & \textbf{2$\times$10${}^{16}$} & 497 & 647 & 0.768 & 6.54 & 1.52 & 5.7 & 0.49 & 0.99 & 19.7 \\ 
\textbf{\textit{}} & \textbf{5$\times$10${}^{16}$} & 539 & 725 & 0.743 & 6.11 & 1.53 & 6.0 & 0.50 & 0.93 & 17.1 \\ 
\textbf{8 nm NbN\newline SNSPD d3\textit{}} & \textbf{0} & 297 & 360 & 0.825 & 8.64 & 1.53 & 4.2 & 0.35 & 1.31 & 50.2 \\ 
\textbf{\textit{}} & \textbf{1$\times$10${}^{16}$} & 382 & 482 & 0.793 & 7.16 & 1.53 & 4.8 & 0.37 & 1.09 & 34.8 \\ 
\end{tabular}
\end{ruledtabular}
\end{table*}
First, we studied the He ion irradiation effects on the NbN film deposited on a DBR substrate. The left axis of the graph in Fig.~\ref{fig:fig1}(b) shows the \textit{T}${}_{c}$ observed for NbN films with different thicknesses (5, 7, 9, and 11 nm) as a function of \textit{F}${}_{i}$ in a linear scale. The right axis of the same graph shows the \textit{F}${}_{i}$ dependence of \textit{$\rho$} only for the 7 nm thick NbN sample. By decreasing the thickness of the NbN film, \textit{T}${}_{c}$ decreases from 9.58 K (at 11 nm) to 5.47 K (at 5 nm). Both \textit{T}${}_{c}$ and \textit{$\rho$} vary monotonically with \textit{F}${}_{i}$. Taking the 7 nm thick film as an example, at the highest \textit{F}${}_{i}$ of 5 $\times$ 10${}^{16}$ ion cm${}^{\mathrm{-}}$${}^{2}$, the ratio of \textit{T}${}_{c}$ and \textit{$\rho$} compared to their initial values are about 0.82 and 1.90, respectively. The experimental data for the \textit{T}${}_{c}$ can be fitted with \textit{T}${}_{c}$ = \textit{a}${}_{1}$log(\textit{F}${}_{i}$) + \textit{b}${}_{1}$, where \textit{a}${}_{1}$ and \textit{b}${}_{1}$ are free parameters. The coefficients of determination, denoted as adjusted R${}^{2}$ in the fittings are all in a range of 0.94$\mathrm{-}$1.0, indicating a good fitting, except for the 5 nm thick sample ($\mathrm{\sim}$0.76), which could be due to the relatively low uniformity of thinner films.

To calculate physical parameters such as the \textit{N}${}_{0}$,\textit{ D}, and Ginzburg--Landau (GL) coherence length (\textit{$\xi$}${}_{GL}$), the upper critical field (\textit{B}${}_{c}$${}_{2}$) of the NbN thin film was recorded as a function of \textit{T} by electrical and magnetic transport measurements, as shown in Fig.~\ref{fig:fig1}(c) for the 7 nm thick films. The dashed lines are the linear fits to the measured data, and the physical parameters were deduced from the dirty limit relation\cite{ref26}, where $B_{c2}\left(0\right)=0.69T_c\frac{{dB}_{c2}}{dT}|_{T\ =\ T_c}$, \textit{$\xi$}${}_{GL}$ = $\sqrt{\frac{{\mathrm{\Phi }}_0}{2\pi B_{\mathrm{c2}}(0)}}$, $N_0$ = \textit{B${}_{c}$}${}_{2}$(0)${\left(\mathrm{0.69}T_c\frac{4ek_B}{\pi }{\rho }_n\right)}^{-1}$, and \textit{D} = $\frac{4k_B}{\pi e}{\left(\frac{{dB}_{c2}}{dT}|_{T\ =\ T_c}\right)}^{-1}$, respectively. In the formulas, \textit{e} is the electron charge, \textit{k}${}_{B}$ is the Boltzmann constant, and \textit{$\mathit{\Phi}$}${}_{0}$ is the flux quantum. The energy gap \textit{$\mathit{\Delta}$}(0) can be calculated using the BCS relation \textit{$\mathit{\Delta}$}(0) = 1.76\textit{k}${}_{B}$\textit{T}${}_{c}$\cite{ref20} and decreases with increasing \textit{F}${}_{i}$. The above results are listed in Table~\ref{tab:TABLE I}. The influence of the irradiation fluence on the physical parameters of the NbN thin film is notable. For example, as \textit{F}${}_{i}$ increased to 5 $\times$ 10${}^{16}$ ion cm${}^{\mathrm{-}}$${}^{2}$, \textit{$\mathit{\Delta}$}(0) and \textit{N}${}_{0}$ decreased by 16\% and 30\%, respectively, and \textit{D }increased by 14\%. According to Eq.~\eqref{1}, assuming \textit{$\tau$${}_{th}$} was unchanged in the ion irradiation process, \textit{E}${}_{min}$ was reduced to about 52\% of its initial value, which indicated that the IDE of the irradiated film could be enhanced.

As shown in Table~\ref{tab:TABLE I}, the superconducting transition width (\textit{$\mathit{\Delta}$T}${}_{c}$, determined by the temperature width where \textit{$\rho$} drops from 90\% to 10\% of \textit{$\rho$}${}_{20 K}$) was in the range from 1.50 to 1.55 K with a small variation of about 0.05 K, indicating that ion irradiation does not cause such damage to the ultrathin film sufficient to impair its uniformity. Therefore, the use of ion irradiation could be an important method for the tuning of the physical properties of thin films. The listed residual resistivity ratio (RRR = \textit{}$\frac{{\rho}_{300K}}{{\rho }_{20K}}$) decreases monotonically with \textit{F}${}_{i}$ because of the increasing disorders. Such reduction was also observed for NbN films with a reduced thickness\cite{ref27}. However, the etching effect in our experiment is negligible, which has been confirmed by the TEM measurements.

Characterizing the optical properties of the irradiated thin films is important because of their applicability for SNSPDs. Figure~\ref{fig:fig1}(d) shows the measured \textit{r} as a function of wavelength (\textit{$\lambda$}) for the 7 nm NbN film at various \textit{F}${}_{i}$s. The variation of \textit{r }caused by different fluences\textit{ }at the target\textit{ $\lambda$} (1550 nm)\textit{ }is less than 1\%. As shown in the inset of Fig.~\ref{fig:fig1}(d), the \textit{r}(\textit{$\lambda$}) curves have a slight red-shift from 1567 nm (at \textit{F}${}_{i}$ = 0) to 1589 nm (at \textit{F}${}_{i}$ = 5 $\times$ 10${}^{16}$ ion cm${}^{\mathrm{-}}$${}^{2}$), taking the position of the minimum \textit{r} (\textit{r}${}_{min}$) as reference. Meanwhile, \textit{r}${}_{min}$ varied from 0\% to 0.3\%. Owing to the broadband ($\mathrm{\sim}$200 nm) and the low reflectance features of the design, the irradiation has a negligible effect on the optical properties of NbN thin films for SNSPD applications.

\begin{figure}
\includegraphics[width=3.6in]{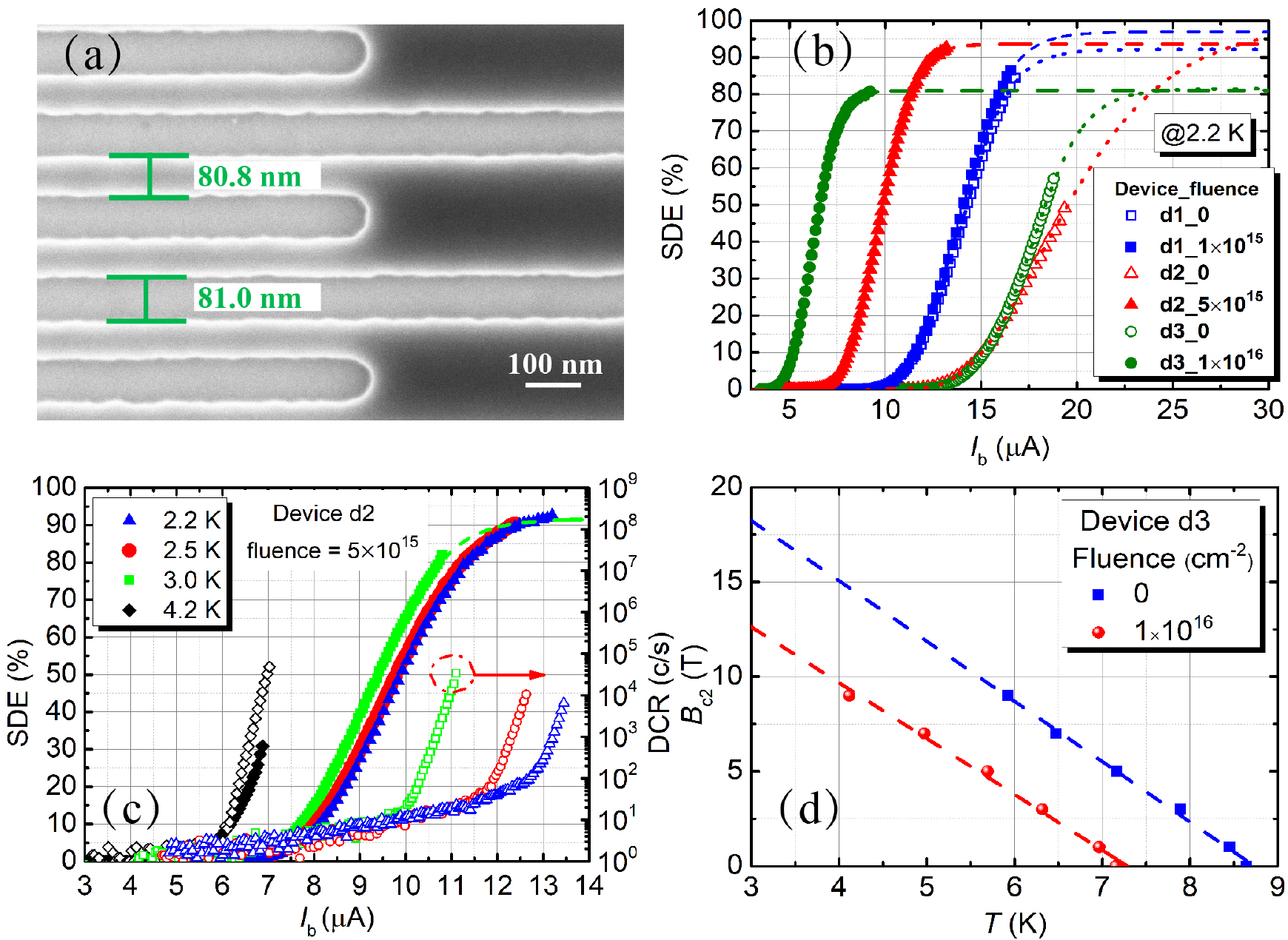}
\caption{\label{fig:fig2}(a) SEM image of 80 nm wide and 160 nm pitch nanowires. (b) SDE as a function of \textit{I${}_{b}$} for the SNSPDs before (open scatters) and after (solid scatters) He ion irradiation. The dashed and dotted lines are the sigmoid function fittings for the data before and after irradiations, respectively. (c) SDE (solid scatters) and DCR (open scatters) of device d2 as a function of \textit{I${}_{b}$} at different temperatures and a fixed \textit{F}${}_{i}$ = 5 $\times$ 10${}^{15}$ ion cm${}^{\mathrm{-}}$${}^{2}$. The green dashed line is the sigmoid function fitting. (d) Temperature dependence of \textit{B}${}_{c2}$ for the device d3 at \textit{F}${}_{i}$ = 0 (squares) and \textit{F}${}_{i}$ = 1 $\times$ 10${}^{16}$ ion cm${}^{\mathrm{-}}$${}^{2}$ (dots). The dashed lines are the linear fittings.}
\end{figure}

Next, the irradiation effects on SNSPDs were investigated for three devices d1, d2, and d3 with a 20 keV He ion irradiation at \textit{F}${}_{i}$ of 1$\times$10${}^{15}$, 5$\times$10${}^{15}$, and 1$\times$10${}^{16}$ ion cm${}^{-2}$, respectively. The detailed information of the SNSPDs used in the experiment is shown in Table~\ref{tab:TABLE II}. The parameters of SNSPDs measured both before and after irradiation are also listed, such as \textit{T}${}_{c}$, switching current (\textit{I}${}_{sw}$), and SDE. The irradiation degraded \textit{T}${}_{c}$ and \textit{I}${}_{sw}$, while \textit{$\mathit{\Delta}$T}${}_{c}$ remained nearly unchanged, consistent with the results for the NbN films. Figure~\ref{fig:fig2}(a) shows a scanning electron microscope (SEM) image of the nanowire.

\begin{table}
\caption{\label{tab:TABLE II}Comparison of the SNSPD parameters before and after 20 keV He ion irradiation, where \textit{d}, \textit{w} and \textit{p} denote the thickness, width and pitch of the nanowre, respectively. }
\begin{ruledtabular}
\begin{tabular}{m{0.8in}m{0.25in}m{0.35in}m{0.25in}m{0.35in}m{0.25in}m{0.35in}} 
\textbf{} & \multicolumn{2}{m{0.8in}}{\centering \textbf{Device d1}} & \multicolumn{2}{m{0.8in}}{\centering \textbf{Device d2}} & \multicolumn{2}{m{0.8in}}{\centering \textbf{Device d3}} \\ 
\textbf{\centering \textit{d} (nm)} & \multicolumn{2}{m{0.5in}}{\centering 7.5} & \multicolumn{2}{m{0.5in}}{\centering 8} & \multicolumn{2}{m{0.5in}}{\centering 8} \\
\textbf{\centering \textit{w}/\textit{p} (nm)} & \multicolumn{2}{m{0.4in}}{\centering {~~~80/160}} & \multicolumn{2}{m{0.4in}}{\centering ~~~80/140} & \multicolumn{2}{m{0.4in}}{\centering ~~~80/140} \\ 
\textbf{\centering \textit{F}${}_{i}$ (cm${}^{\boldsymbol{\mathrm{-}}}$${}^{2}$)} & 0 & 1$\times$10${}^{15}$ & 0 & 5$\times$10${}^{15}$ & 0 & 1$\times$10${}^{16}$ \\ 
\textbf{\centering \textit{T}${}_{c}$ (K)} & 8.3 & 8.1 & 8.6 & 7.5 & 8.6 & 7.2 \\ 
\textbf{\centering \textit{$\boldsymbol{\rho}$}${}_{20 K}$($\mu\Omega\cdot cm$)} & 394 & 435 & 377 & 463 & 360 & 482 \\ 
\textbf{\centering \textit{I}${}_{sw\ }$($\mu$A)} & 16.8 & 16.7 & 19.4 & 12.9 & 18.9 & 9.3 \\ 
\textbf{\centering SDE${}_{max}$(\%)} & 84 & 88 & 49 & 92 & 57 & 81 \\ 
\end{tabular}
\end{ruledtabular}
\end{table}

Figure~\ref{fig:fig2}(b) shows the SDE as a function of\textit{ I}${}_{b}$ for SNSPD before and after irradiation. The maximum SDE (SDE${}_{max}$) significantly increased for the irradiated devices. A weak saturated plateau emerges for the SNSPDs irradiated at \textit{F}${}_{i}$ of 5 $\times$ 10${}^{15}$ and 1 $\times$ 10${}^{16}$ ion cm${}^{\mathrm{-}}$${}^{2}$, indicating the enhancement of their IDEs. An empirical sigmoid function fitting\cite{ref16,ref29} was applied to the experimental data. The asymptotic values (SDE${}_{asy}$) of the fittings imply the hypothetical maximum SDEs for IDEs of 100\%. The SDE of SNSPD can be expressed as SDE = IDE$\cdot$\textit{$\alpha$}$\cdot$OCE\cite{ref9}, where OCE is the optical coupling efficiency. In the asymptotic situation, assuming IDE as unity and the change in \textit{$\alpha$ }induced by irradiation as negligible, SDE mainly depends on OCE. For example, the SDE${}_{asy}$ values for device d3 before and after irradiation are nearly the same, indicating that OCE also does not significantly change, which suggests a good optical coupling. However, owing to the limitation by the accuracy of the optical coupling, there is still a slight difference ($\mathrm{<}$5\%) in the SDE${}_{asy}$ values for two subsequent measurements for device d1 and d2. Nevertheless, the results indicate that there is no significant change in \textit{$\alpha$ }of the nanowire after the irradiation. Thus, the enhancement of SDE is mainly due to the improvement of IDE. Since the IDE is influenced by thickness, width, and constriction in the nanowire, directly comparing the IDE of two different devices is difficult. Moreover, the maximum SDE of about 81\% for device d3 was limited by the contamination on its surface, as found using an optical microscope.

Interestingly, the SDE of device d2 at an \textit{F}${}_{i}$ of 5 $\times$ 10${}^{15}$ ion cm${}^{\mathrm{-}}$${}^{2}$ was significantly enhanced from 49\% to 92\% at 2.2 K. Its SDE is shown in Fig.~\ref{fig:fig2}(c) as a function of \textit{I}${}_{b}$ at various temperatures. With increasing temperature, the saturated platform gradually degrades. Even so, an SDE of 90.4\% was obtained at 2.5 K. Compared to previous results achieved at 2.1 K\cite{ref10}, the operating temperature has been further improved. Even at 3 K, the SDE still exceeds 80\%, making the device attractive for a compact refrigerator compatible with space applications\cite{ref30}.

Figure~\ref{fig:fig2}(d) shows the temperature dependence of \textit{B}${}_{c2}$ for the devices irradiated at \textit{F}${}_{i}$ = 0 and 1 $\times$ 10${}^{16}$ ion cm${}^{\mathrm{-}}$${}^{2}$. The \textit{B}${}_{c2}$ of the nanowire is larger than that of the NbN films shown in Fig.~\ref{fig:fig1}(c) because of the edge barrier effect\cite{ref31}. The physical parameters of device d3 were calculated from the experimental data,as shown in Table~\ref{tab:TABLE I}(lower panel). As expected, \textit{$\mathit{\Delta}$} and \textit{N}${}_{0}$ of the irradiated samples decreased, and \textit{D} remained nearly unchanged, compared with the original values. \textit{N}${}_{0}$ is larger in the nanowire than in the thin film because a thicker layer was used for SNSPD. The estimated \textit{E}${}_{min}$ for the nanowire irradiated at \textit{F}${}_{i}$ = 1 $\times$ 10${}^{16}$ ion cm${}^{\mathrm{-}}$${}^{2}$ was reduced to nearly 50\% of its initial value, calculated from Eq.~\eqref{1}, which explains the enhancement of the IDE.

The reduction in \textit{$\mathit{\Delta}$} as well as \textit{N}${}_{0}$ and increasing in \textit{$\rho$}${}_{n}$ were observed in our experiments with He ion irradiations. The mechanism could be explained with the displacement of atoms\cite{ref1}, which generated vacancies in the NbN thin film then influenced its specific charge transport\cite{ref3}. Furthermore, the ultrathin polycrystalline NbN films in this paper demonstrated a stronger reduction (up to 23\% for 5 nm) in \textit{T}${}_{c}$ than that thicker polycrystalline films\cite{ref3}(6\%--7\%, 45--135 nm thick) irradiated with 200 keV Ar ions. This phenomenon indicates that thinner films are more easily influenced by irradiation. 
As regards the devices, a direct method to evaluate the influence of He ion irradiation on SNSPDs was introduced. The IDE enhancement on irradiated SNSPDs was achieved owing to the ion-induced reduction of \textit{$\mathit{\Delta}$ }and \textit{N}${}_{0}$. The irradiation did not degrade \textit{$\mathit{\Delta}$T}${}_{c}$ of the devices, enabled us to continuously tune the performance of the fabricated devices. This method has a double advantage. On one hand, it is a post-processing technology, compatible with the micro-electronics industry. On the other hand, it can relax the fabrication precision on the film thickness as well as the width of the nanowire and also extend the spectral responsivity of the device to mid- or even far-infrared range by reducing \textit{$\mathit{\Delta}$}, similar to what WSi devices did\cite{ref32}. In the future, there are a few interesting works to explore. 
Such as, investigations on the yield improvement using either pre-irradiated NbN/DBR wafers or post-irradiated SNSPDs would be interested for the applications requiring large amount of SNSPDs\cite{ref33}.

In conclusion, we reported a flexible method to tune the physical properties of superconducting polycrystalline NbN ultrathin films and SNSPDs using He ion irradiation. The irradiation induced vacancies in NbN films led to a monotonic decrease in superconducting energy gap and the electron density of states at the Fermi level as He ion fluence increased. The influence on the optical absorptance of the irradiated NbN film can be neglected when fabricating SNSPDs. Using He irradiation directly on the SNSPDs, IDE and SDE were significantly improved. The SDE of an irradiated device at a He fluence of 5 $\times$ 10${}^{15}$ ion cm${}^{\mathrm{-}}$${}^{2}$ was significantly enhanced from 49\% to 92\% at 2.2 K, and over 90\% at 2.5 K, for a telecom wavelength of 1550 nm. The He ion irradiation is a promising tool for enhancement of SNSPD's properties, such as relaxing the process requirements for film thickness and linewidth control, IDE enhancement for longer wavelength, and yield improvement. Ion irradiation may also be applied in the fabrication of other superconducting devices.

This work is supported by National Key R\&D Program of China (2017YFA0304000), NSFC (11622545, and U1732268), and STCSM (16JC1400402).


\begin{thebibliography}{99}
\bibitem{ref1}M. Nastasi, J. Mayer, and J. K. Hirvonen, Ion-Solid Interactions: Fundamentals and Applications. (Cambridge University Press, Cambridge, 1996).
\bibitem{ref2}A. Herklotz, S. F. Rus, and T. Z. Ward,  Nano Lett. \textbf {16}, 1782 (2016);  X. Ou, Y. Shuai, W. B. Luo, P. F. Siles, R. Koegler, J. Fiedler, H. Reuther, S. Q. Zhou, R. Huebner, S. Facsko, M. Helm, T. Mikolajick, O. G. Schmidt, and H. Schmidt,  Acs Appl. Mater. Inter. \textbf {5}, 12764 (2013);  Q. Jia, J. Grenzer, H. B. He, W. Anwand, Y. D. Ji, Y. Yuan, K. Huang, T. G. You, W. J. Yu, W. Ren, X. Z. Chen, M. K. Liu, S. Facsko, X. Wang, and X. Ou,  Adv. Mater. Interfaces \textbf {5}, 1701268 (2018).
\bibitem{ref3}J. Y. Juang, D. A. Rudman, J. Talvacchio, and R. B. van Dover,  Phys. Rev. B \textbf {38}, 2354 (1988).
\bibitem{ref4}Y. Bugoslavsky, L. F. Cohen, G. K. Perkins, M. Polichetti, T. J. Tate, R. Gwilliam, and A. D. Caplin,  Nature \textbf {411}, 561 (2001);  M. Leroux, K. J. Kihlstrom, S. Holleis, M. W. Rupich, S. Sathyamurthy, S. Fleshler, H. P. Sheng, D. J. Miller, S. Eley, L. Civale, A. Kayani, P. M. Niraula, U. Welp, and W.-K. Kwok,  Appl. Phys. Lett. \textbf {107}, 192601 (2015).
\bibitem{ref5}C. Tarantini, K. Iida, N. Sumiya, M. Chihara, T. Hatano, H. Ikuta, R. K. Singh, N. Newman, and D. C. Larbalestier,  Supercond. Sci. Tech. \textbf {31}, 034002 (2018).
\bibitem{ref6}M. Putti, M. Affronte, C. Ferdeghini, P. Manfrinetti, C. Tarantini, and E. Lehmann,  Phys. Rev. Lett. \textbf {96}, 077003 (2006).
\bibitem{ref7}S. A. Cybart, E. Y. Cho, T. J. Wong, B. H. Wehlin, M. K. Ma, C. Huynh, and R. C. Dynes,  Nat. Nanotechnol. \textbf {10}, 598 (2015).
\bibitem{ref8}D. Dew-Hughes and R. Jones, Appl. Phys. Lett. \textbf {36}, 856 (1980).
\bibitem{ref9}F. Marsili, V. B. Verma, J. A. Stern, S. Harrington, A. E. Lita, T. Gerrits, I. Vayshenker, B. Baek, M. D. Shaw, R. P. Mirin, and S. W. Nam,  Nat. Photon. \textbf {7}, 210 (2013).
\bibitem{ref10}W. J. Zhang, L. X. You, H. Li, J. Huang, C. L. Lv, L. Zhang, X. Y. Liu, J. J. Wu, Z. Wang, and X. M. Xie,  Sci. China Phys. Mech. Astron. \textbf {60}, 120314 (2017).
\bibitem{ref11} W. J. Zhang, X. Y. Yang, H. Li, L. X. You, C. L. Lv, L. Zhang, C. J. Zhang, X. Y. Liu, Z. Wang, and X. M. Xie,  Supercond. Sci. Tech. \textbf {31}, 035012 (2018).
\bibitem{ref12}J. J. Wu, L. X. You, S. J. Chen, H. Li, Y. H. He, C. L. Lv, Z. Wang, and X. M. Xie,  Appl. Opt. \textbf {56}, 2195 (2017);  I. E. Zadeh, J. W. N. Los, R. B. M. Gourgues, V. Steinmetz, G. Bulgarini, S. M. Dobrovolskiy, V. Zwiller, and S. N. Dorenbos,  APL Photon. \textbf {2}, 111301 (2017).
\bibitem{ref13}Q. -C. Sun, Y. -L. Mao, S. -J. Chen, W. Zhang, Y. -F. Jiang, Y. -B. Zhang, W. -J. Zhang, S. Miki, T. Yamashita, H. Terai, X. Jiang, T. -Y. Chen, L. -X. You, X. -F. Chen, Z. Wang, J. -Y. Fan, Q. Zhang, and J. -W. Pan,  Nat. Photon. \textbf {10}, 671 (2016).
\bibitem{ref14}Y. He, X. Ding, Z. E. Su, H. L. Huang, J. Qin, C. Wang, S. Unsleber, C. Chen, H. Wang, Y. M. He, X. L. Wang, W. J. Zhang, S. J. Chen, C. Schneider, M. Kamp, L. X. You, Z. Wang, S. H\"{o}fling, C. -Y. Lu, and J. -W. Pan,  Phys. Rev. Lett. \textbf {118}, 190501 (2017).
\bibitem{ref15}B. S. Robinson, A. J. Kerman, E. Dauler, R. J. Barron, D. O. Caplan, M. L. Stevens, J. J. Carney, S. A. Hamilton, J. K. Yang, and K. K. Berggren,  Opt. Lett. \textbf {31}, 444 (2006).
\bibitem{ref16}T. Yamashita, S. Miki, H. Terai, and Z. Wang,  Opt. Express \textbf {21}, 27177 (2013).
\bibitem{ref17}G. N. Gol'tsman, O. Okunev, G. Chulkova, A. Lipatov, A. Semenov, K. Smirnov, B. Voronov, A. Dzardanov, C. Williams, and R. Sobolewski,  Appl. Phys. Lett. \textbf {79}, 705 (2001).
\bibitem{ref18}Y. P. Korneeva, M. Y. Mikhailov, Y. P. Pershin , N. N. Manova, A. V. Divochiy, Y. B. Vakhtomin, A. A. Korneev, K. V. Smirnov, A. G. Sivakov, A. Y. Devizenko, and G. N. Goltsman,  Supercond. Sci. Tech. \textbf {27}, 095012 (2014).
\bibitem{ref19}V. B. Verma, B. Korzh, F. Bussi\`{e}res, R. D. Horansky, S. D. Dyer, A. E. Lita, I. Vayshenker, F. Marsili, M. D. Shaw, H. Zbinden, R. P. Mirin, and S. W. Nam,  Opt. Express \textbf {23}, 33792 (2015).
\bibitem{ref20}A. Engel, A. Aeschbacher, K. Inderbitzin, A. Schilling, K. Il'in, M. Hofherr, M. Siegel, A. Semenov, and H.-W. H\"{u}bers,  Appl. Phys. Lett. \textbf {100}, 062601 (2012).
\bibitem{ref21}Y. Ivry, J. J. Surick, M. Barzilay, C.-S. Kim, F. Najafi, E. Kalfon-Cohen, A. D. Dane, and K. K. Berggren,  Nanotechnology \textbf {28}, 435205 (2017).
\bibitem{ref22}D. Henrich, S. D\"{o}rner, M. Hofherr, K. Il'in, A. Semenov, E. Heintze, M. Scheffler, M. Dressel, and M. Siegel,  J. Appl. Phys. \textbf {112}, 074511 (2012);  A. E. Dane, A. N. McCaughan, D. Zhu, Q. Y. Zhao, C.-S. Kim, N. Calandri, A. Agarwal, F. Bellei, and K. K. Berggren,  Appl. Phys. Lett. \textbf {111}, 122601 (2017).
\bibitem{ref23}A. Semenov, A. Engel, H. W. Hubers, K. Il'in, and M. Siegel,  Eur. Phys. J. B \textbf {47}, 495 (2005).
\bibitem{ref24}F. Marsili, F. Najafi, E. Dauler, F. Bellei, X. L. Hu, M. Csete, R. J. Molnar, and K. K. Berggren,  Nano lett. \textbf {11}, 2048 (2011).
\bibitem{ref25}James F. Ziegler, M. D. Ziegler, and J. P. Biersack,  Nucl. Instrum. Methods Phys. Res., Sect. B \textbf {286}, 1818 (2010).
\bibitem{ref26}N. R. Werthamer, E. Helfand, and P. C. Hohenberg,  Phys. Rev. \textbf {1472}, 295 (1966).
\bibitem{ref27}S. Konstantin, D. Alexander, V. Yury, M. Pavel, Z. Philipp, A. Andrey, and S. Vitaliy,  Supercond. Sci. Tech. \textbf {31}, 035011 (2018).
\bibitem{ref29}A. G. Kozorezov, C. Lambert, F. Marsili, M. J. Stevens, V. B. Verma, J. P. Allmaras, M. D. Shaw, R. P. Mirin, and S. W. Nam,  Phys. Rev. B \textbf {96}, 054507 (2017).
\bibitem{ref30}L. X. You, J. Quan, Y. Wang, Y. X. Ma, X. Y. Yang, Y. J. Liu, H. Li, J. G. Li, J. Wang, J. T. Liang, Z. Wang, and X. M. Xie,  Opt. Express \textbf {26}, 2965 (2018).
\bibitem{ref31}G. M. Maksimova,  Phys. Solid State \textbf {40}, 1607 (1998).
\bibitem{ref32}L. Chen, D. Schwarzer, V. B. Verma, M. J. Stevens, F. Marsili, R. P. Mirin, S. W. Nam, and A. M. Wodtke,  Accounts Chem. Res. \textbf {50}, 1400 (2017).
\bibitem{ref33}H. Wang, W. Li, X. Jiang, Y. M. He, Y. H. Li, X. Ding, M. C. Chen, J. Qin, C. Z. Peng, C. Schneider, M. Kamp, W. J. Zhang, H. Li, L. X. You, Z. Wang, J. \ P Dowling, S. H\"{o}fling, C-Y Lu, and J-W Pan,  Phys. Rev. Lett. \textbf {120}, 230502 (2018).

\end{thebibliography}
\end{document}